\documentclass[review]{elsarticle}
\usepackage{amssymb}

\newcommand{\clV}{\mathcal{ V}}

\newcommand{\clM}{\mathcal{ M}}

\newcommand{\clL}{{\cal L}}

\newcommand{\clE}{\mathcal{E}}

\newcommand{\prt}{\partial}

\newcommand{\rgl}{\rangle}
\newcommand{\lgl}{\langle}
\newcommand{\ep}{\epsilon}
\newcommand{\vep}{\varepsilon}
\newcommand{\be}{\begin{equation}}
\newcommand{\ee}{\end{equation}}
\newcommand{\bea}{\begin{eqnarray}}
\newcommand{\eea}{\end{eqnarray}}

\begin{document}

\begin{frontmatter}

\title{Electrostatics in Fractal Geometry: Fractional Calculus Approach}

\author[lab1,lab2]{Emmanuel Baskin}
\ead{baskin@tx.technion.ic.il}
\author[lab1]{Alexander Iomin\corref{cor}}
\ead{iomin@physics.technion.ac.il}

\cortext[cor]{Corresponding author}

\address[lab1]{Department of Physics and Solid State Institute,
Technion, Haifa, 32000, Israel}

\address[lab2]{ H4 Energy Solution Ltd, 4 Pakris st., Rehovot
76702, Israel }

\begin{abstract}

The electrostatics properties of composite materials with fractal
geometry are studied in the framework of fractional calculus.
An electric field in a composite dielectric with a fractal charge
distribution is obtained in the spherical symmetry case. The
method is based on the splitting of a composite volume into a fractal
volume $V_d\sim r^d$ with the fractal dimension $d$ and a
complementary host volume $V_h=V_3-V_d$. Integrations over these
fractal volumes correspond to the convolution integrals that
eventually lead to the employment of the fractional
integro-differentiation.

%%\pacs{05.45.Df, 41.20.Cv, 41.20.-q}

\end{abstract}


\end{frontmatter}

\section{Introduction}
The electrostatics of composite materials is a ubiquitous and
fundamental phenomenon  with a variety of applications in
physical, electro-chemical, geophysical, and biological systems,
see \textit{e.g.} \cite{mandelbrot,gouyet}. The investigation of
the electric properties of composite materials has been a
long-lasting task, starting, \textit{e.g.}, from a calculation of
the permittivity of a dispersed mixture \cite{LLVIII} and
continuing with the modern investigations of nanosystems, for
example, of an enhanced scattering of electric fields in
nanosystems \cite{b8,st1,st1b}. The concept of fractals and
fractal geometry is widely employed in these studies, and leads to
essential progress in our understanding of the physical properties
of composites.  A wide class of the physical applications of
fractal structures has been reviewed well, see \textit{e.g.},
\cite{mandelbrot,gouyet,west90,OLE93}. In its turn, the fractal
concept makes it possible to involve the application of the
powerful methods of fractional integro-differentiation. An
illuminating example of this relation is fractional diffusion on a
comb \cite{em1,benAv,bi2004}, where the relation between the comb
geometry and non-Markovian transport is straightforwardly
established in the fractional Fokker-Planck equation
\cite{benAv,bi2004,ib2005}. As obtained in \cite{mandelvanness}
(see also \cite{mandelbrot}  chapter 39), the mathematical
formulation of the transform of the Brown line-to-line function
$B(t)$ (continuous, but not differentiable) into fractional
diffusion, described by fractional line-to-line function
$B_{\alpha}(t)$ with $0<\alpha<2$, is expressed by means of the
fractional integration \footnote{Note that fractional Brown motion
was originally introduced by Kolmogorov in \cite{kolmogorov} and
was later rediscovered and explored in greater detail by
Mandelbrot and Van Ness in \cite{mandelvanness}.}
\be\label{mandel}   %
B_{\alpha}(t)=\frac{1}{\Gamma(\alpha)}
\int_{-\infty}^t(t-s)^{\alpha-1}dB(s)\,
. \ee   %
 The concept of the integration and
differentiation of non-integer orders already arises from the
works of Leibniz, Liouville, and Riemann,  see \textit{e.g.},
\cite{podlubny,oldham}. Its application is related to random
processes with power law distributions. This corresponds to the
absence of characteristic average values for processes that
exhibit many scales \cite{klafter,shlesinger}.

By analogy with Eq. (\ref{mandel}), a link between  fractal
geometry and fractional integro-differentiation \cite{NIG92} is
constituted in the
procedure of averaging an extensive physical value that is
expressed by means of a smooth function over a Cantor set, which
leads to fractional integration. In its eventual form, it has been
presented in Ref. \cite{NIG98}. The main idea of filtering or
embedding a matter inside a fractal is the construction of a
convolution integral
\be\label{fpe1}    %
F(x)=W(x)\star f(x)=\int_0^xW(x-y)f(y)dy\, , \ee  %
where the function $W(x)$ obeys the scaling relation
$W(x)=\frac{1}{a}W(bx)$ with a solution $W(x)=x^{\alpha}A(x)$.
Here $\alpha=\frac{\ln a}{\ln b}$ and $A(x)=A(bx)$ is a
log-periodic function with a period $\ln b$. When this scaling
corresponds to a Cantor set with $a=2$, and $A(x)$ is defined
explicitly, one performs averaging over this period \cite{NIG98}
and obtains Eq. (\ref{fpe1}) in the form of a fractional integral
\be\label{fpe3}   %
\lgl{F(x)}\rgl=\frac{\clV(\alpha)}{\Gamma(\alpha)}\Big(x^{\alpha-1}\star
f(x)\Big)\equiv \clV(\alpha){}_0I_x^{\alpha}f(x)\, ,  \ee    %
where $\clV(\alpha)=\frac{2^{-1+\alpha/2}}{\ln 2}$ and
$\Gamma(\alpha)$ is a gamma function, while ${}_0I_x^{\alpha}$
designates the fractional integral of the order of $\alpha$ in the
range from $0$ to $x$:
\be\label{fpe4}   %
{}_0I_x^{\alpha}f(x)=\frac{1}{\Gamma(\alpha)}
\int_0^x(x-x')^{\alpha-1}f(x')dx'\,
.  \ee   %
Analogously one can define the fractional derivative as the
inversion operation: ${}_0D_x^{\alpha}={}_0I_x^{-\alpha}$.
Extended reviews on fractional calculus can be found in
\cite{podlubny,oldham,SKM}, for example. See  Appendix A.

This mathematical construction is relevant to the study of
electrostatics of real composite structures. The main objective of
the present research is to solve a standard electrostatic problem,
namely, the derivation of the electric field in a composite dielectric
with a fractal charge distribution. This problem was first
considered in \cite{tar2005}, in a slightly different mathematical
formulation, where integration was carried out over the fractal
with fractal dimension $\alpha$ and infinitesimal fractal volume
$dr^{\alpha}$ inside a homogeneous dielectric.

\section{Fractional Distribution of Charges}

First, we demonstrate this fractal property of embedding a matter
for electrostatics by implementing it in Maxwell's integral
equation for the electric field $\mathbf{E}$. We formulate the
problem in terms of a characteristic function. Let any charge
distribution on a fractal be $F_{\alpha}(\mathbf{r})$, where
$\mathbf{r}=(x,y,z)$ is described by the characteristic function
$\chi(\mathbf{r})$, such that $\chi(\mathbf{r})=1$ for
$\mathbf{r}\in F_{\alpha}(\mathbf{r})$ and $\chi(\mathbf{r})=0$
for $\mathbf{r}\notin F_{\alpha}(\mathbf{r})$. Therefore, the
fractal filtering of Eq. (\ref{fpe3}) now reads
\be\label{rfc_1}   %
\int_0^x\chi(x')f(x')dx'\Rightarrow\frac{1}{\Gamma(\alpha)}
\int_0^x(x-x')^{\alpha-1}f(x')dx'\equiv {}_0I_x^{\alpha}f(x)\, .\ee  %
We obtain the same integration for the $y$ and $z$ coordinates.

The Gauss theorem in Cartesian's coordinates reads
\be\label{fpe5_ME}   %
\int\nabla\cdot\mathbf{E}\, dxdydz\equiv
 {}_0I_x^{1}{}_0I_y^{1}{}_0I_z^{1} \nabla\cdot\mathbf{E}=Q\,  ,  \ee  %
where an electric charge $Q$ with a fractional distribution is
embedded in a three-dimensional volume. Without restricting the
generality, we consider the fractal charge in the first octant of
the Euclidian space. Therefore, we have from Eq. (\ref{rfc_1})
\be\label{fpe6}  %
Q\equiv
Q(x,y,z)={}_0I_x^{1}\chi(x){}_0I_y^{1}
\chi(y){}_0I_z^{1}\chi(z)\rho(x,y,z)
\Rightarrow{}_0I_x^{\alpha}{}_0I_y^{\alpha}{}_0I_z^{\alpha}\rho(x,y,z)
\, ,  \ee     %
where $\rho(x,y,z)$ is the charge density, and the constant
multiplier $\clV(\alpha)$ is inserted in $\rho$; for an alternate
definition see also \cite{tar2005}. The charge in Eq. (\ref{fpe6})
corresponds to a fractal structure that is the direct product of
the Cantor sets defined in Eq. (\ref{fpe3}), namely $\lgl
F_{\alpha}(x)\rgl\times\lgl F_{\alpha}(y)\rgl\times\lgl
F_{\alpha}(z)\rgl$.

To proceed, we obtain the correct dimensions of both the charge
and the electric field, respectively. To this end it is convenient
to work in the dimensionless space variables
$\frac{x}{l_0}\rightarrow x, ~\frac{y}{l_0}\rightarrow
y,~\frac{z}{l_0}\rightarrow z$, and the properly defined fractal
charge reads from Eq. (\ref{fpe6})
\be\label{fpe6_a}   %
Q(x,y,z)=
{}_0I_x^{\alpha}{}_0I_y^{\alpha}{}_0I_z^{\alpha}l_0^3\rho(x,y,z)\,
.  \ee  %
The conventional integral in Eq. (\ref{fpe5_ME}) yields the
dimensionless integration
${}_0I_x^{1}{}_0I_y^{1}{}_0I_z^{1}\rightarrow
l_0^3{}_0I_x^{1}{}_0I_y^{1}{}_0I_z^{1}$ and dimensionless
differentiation $\nabla \rightarrow\frac{1}{l_0}\nabla$. Here
$l_0$ is the characteristic size of a physical fractal, which is
always finite. For example, it can be the minimal scale of
self-similarity of a physical fractal. In what follows we always
bear in mind that the space variables are dimensionless, and the
charge density is properly scaled such that fractal charges and
electric fields have correct dimensions.

When the charge density is a constant function, the fractional
convolution integrals become a simple integration over a
dimensionless  volume $(xyz)^{\alpha}/\Gamma^3(\alpha+1)$.
Comparison of  Eqs. (\ref{fpe5_ME}) and (\ref{fpe6})  yields
Maxwell's equation in dimensionless space coordinates
\be\label{fpe7}   %
\nabla\cdot\mathbf{E}=\frac{\prt^3Q}{\prt x\prt y\prt
z}=\frac{(xyz)^{\alpha-1}}{\Gamma^3(\alpha)}l_0\rho\,
.   \ee    %
It has a solution
\be\label{fpe8} %
\mathbf{E}=\frac{l_0\rho}{\alpha\Gamma^3(\alpha)}
\Big[\hat{\mathbf{x}}x^{\alpha}
(yz)^{\alpha-1}+\hat{\mathbf{y}}y^{\alpha}(xz)^{\alpha-1}+
\hat{\mathbf{z}}z^{\alpha}
(xy)^{\alpha-1}\Big] \, . \ee  %
When the charge density $\rho(x,y,z)$ is not a constant, the
integrands in Eqs. (\ref{fpe5_ME}) and (\ref{fpe6}) cannot be
straightforwardly compared to each other. Therefore, we apply the
fractional differential ${}_0D_x^{\alpha}$ to both sides of
Eq. (\ref{fpe5_ME}). Taking into account that
${}_0D_x^{\alpha}{}_0I_x^{\alpha}f(x)=f(x)$ and
${}_0D_x^{\alpha}{}_0I_x^{\beta}f(x)={}_0D_x^{\alpha-\beta}f(x) $
(see Appendix, Eq. (\ref{Ap6})), one obtains Maxwell's
equation in form either
\be\label{fpe9}   %
{}_0I_x^{1-\alpha}{}_0I_y^{1-\alpha}{}_0I_z^{1-\alpha}
\nabla\cdot\mathbf{E}(x,y,z)=\rho(x,y,z)\, , \ee   %
or
\be\label{fpe10}   %
\nabla\cdot\mathbf{E}(x,y,z)=
{}_0D_x^{1-\alpha}{}_0D_y^{1-\alpha}{}_0D_z^{1-\alpha}\rho(x,y,z)\,
. \ee   %
For a constant $\rho$ it reduces to Eq. (\ref{fpe7}).

\section{Example: Electric Field of an Infinite Fractal String}

Now we consider an example where the filtering by means of the
convolution integral in Eqs. (\ref{fpe3}) and (\ref{fpe4}) can be
considered as a superposition of electric fields, that is the
fractal filtering of the electric field. Let us find an electric
field of a charged fractal string with the fractal dimension
$0<\nu<1$ and a constant linear charge density  $\rho_1$. We use
cylindrical symmetry and describe the problem in terms of the
string coordinate $z$ and the distance from the string $r$. The
$E_r$ component of the electric field at point $(r,z)$ of an
infinitesimally small charge with coordinate $z'$ is
$dE_r(r,z)=\frac{\rho_1rdz}{l_0[r^2+(z-z')^2]^{\frac{3}{2}}}$,
where we used that the linear charge density is scaled
$\frac{\rho_1}{l_0}$. Therefore, the complete electric field at
this point is a superposition of the charges along the fractal
string. This yields the fractional integration
\be\label{string_1}  %
E_r(r,z)=\frac{r\rho_1}{l_0\Gamma(\nu)}
\Big[\int_{-\infty}^z
\frac{(z-z')^{\nu-1}dz'}{[r^2+(z-z')^2]^{\frac{3}{2}}}+
\int_z^{\infty}
\frac{(z'-z)^{\nu-1}dz'}{[r^2+(z-z')^2]^{\frac{3}{2}}}\Big]\,
. \ee  %
Performing the variable change $y=(z-z')/r$, one obtains for the
electric field
\be\label{string}  %
E_r(r,z)\equiv E_r(r)=\frac{2A(\nu)\rho_1}{l_0r^{2-\nu}}\, , \ee
where $A(\nu)=\frac{1}{\Gamma(\nu)}\int_0^{\infty}
\frac{y^{\nu-1}dy}{[1+y^2]^{\frac{3}{2}}}$. When $\nu=1$, this
yields a well known result for the electric field of an infinitely
long line charge. In dimension variables the electric field reads
$E_r(r)=2A(\nu)\Big(\frac{l_0}{r}\Big)^{1-\nu}\frac{\rho_1}{r}$,
\textit{i.e.}, the electric field of a fractal string decreases
with the distance $r$ $(r/l_0)^{1-\nu}$ times faster than one
of the continuous counterparts.

\section{A Case of Spherical Symmetry}

Now we obtain an expression for the fractal charge in
spherical coordinates. We define a charge on a random fractal,
which has a homogeneous distribution in a three-dimensional medium
with charge density $\rho_0$. A fractal mass inside a ball of
radius $r$ is $\clM(r)\sim r^d$. As shown in Sec. 2, the
convolution in Eq. (\ref{fpe6}) determines the map of the charge into
the fractal volume. In Cartesian's coordinates it corresponds to
the fractional Riemann-Liouville integral with elementary
fractional volume \cite{SKM,tar2005}
\[dV_d=\frac{|xyz|^{\frac{d}{3}-1}}{\Gamma^3(d/3)}dxdydz\, .\]
In the spherical coordinates, which correspond to the Reisz
definition of the fractional integral, the elementary fractional
volume is
\[dV_d=\frac{2^{3-d}\Gamma(3/2)}{\Gamma(d/2)}|
\mathbf{r}|^{d-3}r^2dr\sin\theta
d\theta d\phi\, .\]  %
In the spherically symmetrical case, the convolution integral
(\ref{fpe6}) takes a simple form. Hence, the random fractal can be
considered as a direct product of the two-dimensional sphere $S_2$
and a Cantor set $F_{\alpha}$ of the dimension $\alpha$, such that
the fractal dimension of $S_2\otimes F_{\alpha}$ is $d=2+\alpha$
\cite{falconer}. Therefore, the mapping of the charge inside the
fractal is determined by the convolution
\be\label{fifr1} %
Q(r)=
\frac{1}{\Gamma(\alpha)}\int_0^r(r-r')^{\alpha-1}\rho_0(r')r'^2dr'\, ,
\ee   %
where all constants are taken inside the fractional charge density
$\rho$. In what follows the charge density $\rho_0$ is a constant
value. This choice immediately yields an expression for the charge
from Eq. (\ref{fifr1})
\be\label{fifr1_a}  %
Q(r)=\frac{2\rho_0}{\Gamma(\alpha+3)} r^d\, ,\ee %
that correlates with the fractional mass $\clM(r)$.  This also
yields the average fractal charge density of the entire composite
\be\label{fifr_3_4}  %
\rho(r)\sim\frac{3Q(r)}{4\pi r^3}\sim\rho_0r^{d-3}\, .
\ee %
This expression yields the corect result for the electric field
inside the fractal dielectric
\be\label{fifr3_5} %
E(R)=\int\frac{\rho(R)dV}{R^2}\sim\rho_0R^{d-2} \ee  %
when $R<r$ \cite{tar2005}. Note also that for the electric field
outside the charged fractal, when $R>r$, we have for the
average fractal charge density
$\rho(R)\rightarrow\rho(r')\Theta(r'-r)$. This
yields for the electric field $E(R)=\frac{Q(r)}{4\pi R^2}$. Here
$\Theta(z)$ is the Heaviside step function.

\section{Electric Field in a Fractal Dielectric Composite}

The situation with respect to permittivity is different and
needs special care. Let us consider the Maxwell equation
(\ref{fpe5_ME}) inside a volume that contains a fractal
dielectric. We take into account
that the electric field is in a random fractal of a volume
$V_d\sim r^d$ with permittivity $\vep_1$ and in a free space
of a complementary host volume $V_h=V_3-V_d$ with permittivity
$\vep_2$. We suppose that the random fractal is a direct product
of the two-dimensional sphere $S_2$ and a Cantor set $F_{\alpha}$
of the dimension $\alpha$, such that the fractal dimension of
$S_2\otimes F_{\alpha}$ is $d=2+\alpha$.  We also suppose the
spherical symmetry,
$\nabla_{\theta}E_{\theta}=\nabla_{\phi}E_{\phi}=0$. The electric
displacement $\mathbf{D}(r)=\vep(r)\mathbf{E}(r)$ is not
differentiable, since the permittivity is a discontinuous function
on the fractal:
\be\label{fifr5}  %
\vep(r)=\left\{
\begin{array}{l}
\vep_1\, , \\
\vep_2\, ,
\end{array}
\begin{array}{l}
\mbox{if $\quad r\in V_d$ } \\
\mbox{if $\quad r\in V_h$}\, .
\end{array}
\right. \ee  %
The permittivity of a two-phase composite in Eq. (\ref{fifr5}) can
be defined by the characteristic function
$\vep(r)=\vep_1\chi(r)+\vep_2[1-\chi(r)]$. The Maxwell equation
for the $\hat{\mathbf{r}}$ component reads
\be\label{fifr6_me}  %
(\vep_1-\vep_2)\chi(r)\nabla_rE_r+
\vep_2\nabla_rE_r+(\vep_1-\vep_2)E_r\nabla_r\chi(r)
=\rho_0\, .\ee%
The last term in the l.h.s. corresponds to a charge, which is due to
the polarization of the fractal and to the discontinuity of the
electric field on the fractal interface \cite{LLVIII}. Now we use the
Gauss theorem, which for the chosen geometry reads
\be\label{fifr6_GT} %
Q(r)=4\pi\int_0^r\nabla_rD_r(r')r'^2dr'\equiv
4\pi{}_0I_r^1r'^2\nabla_rD_r(r)\, . \ee %
Taking into account that
${}_0I_r^1\chi(r)r^2f(r)\rightarrow{}_0I_r^{\alpha}r^2f(r)$, we
obtain from Eqs. (\ref{fifr6_me}) and (\ref{fifr6_GT})
\be\label{fifr6_a}   %
\frac{1}{4\pi}Q(r)=(\vep_1-\vep_2)\,{}_0I_r^{\alpha}r^2\nabla_rE_r(r)
+\vep_2\,{}_0I_r^1r^2\nabla_rE_r(r)+
(\vep_1-\vep_2)\,{}_0I_r^1E_r(r)r^2\nabla_r\chi(r)\, .\ee  %
The last term in Eq. (\ref{fifr6_a}) can be written in the form
(see Appendix B) ${}_0I_r^{\alpha}E_r(r)r^2$. Finally, taking into
account Eq. (\ref{fifr1_a}), we obtain the integral Maxwell's
equation for the electric field in the form
\be\label{fifr6_b}  %
(\vep_1-\vep_2)\,{}_0I_r^{\alpha}r^2\nabla_rE_r(r)
+\vep_2\,{}_0I_r^1r^2\nabla_rE_r(r)+
\frac{(\vep_1-\vep_2)^2}{\vep_2}{}_0I_r^{\alpha}E_r(r)r^2=
\frac{2\rho_0}{\Gamma(\alpha+3)}r^{2+\alpha}\, . \ee %
We take into account that
$r^2\nabla_rE_r(r)=\prt_r[r^2E_r(r)]\equiv \prt_rG(r)$, where we
introduce a new function, $G(r)=r^2E_r(r)$. We also introduce a
dimensionless parameter
\be\label{fifr6_c}  %
\ep=\frac{\vep_1-\vep_2}{\vep_2}\, . \ee  %
Applying the fractional derivative ${}_0D_r^{\alpha}$ to both
sides of Eq. (\ref{fifr6_b}) and taking into account that
${}_0D_r^{q}r^p=\frac{\Gamma(p+1)}{\Gamma(p+1-q)}r^{q-d}$ one
obtains
\be\label{fifr6}    %
\ep\prt_rG(r)+{}_0I_r^{1-\alpha}\prt_rG(r)+
\ep^2G(r)=\frac{r^2\rho_0}{\vep_2} \, . \ee %
Here
\be\label{fifr7}   %
{}_0I_r^{1-\alpha}\prt_rG(r)=
\frac{1}{\Gamma(1-\alpha)}\int_0^r(r-r')^{-\alpha}\prt_{r'}G(r')dr'\, .
\ee   %
It is worth noting that for a homogeneous medium with a fractal
distribution of charges, when $\vep_1=\vep_2$, Eq. (\ref{fifr6})
reduces to Eq. (\ref{fpe7}) in the spherical symmetry.

\subsection{Solution by Expansion}

Eq. (\ref{fifr6}) is the linear fractional integro-differential
equation that can be solved by the Laplace transform.
Defining the Laplace transform
$\hat{\clL}[G(r)]=\tilde{G}(\lambda)$, one obtains the solution of
Eq. (\ref{fifr7}) in the form
\begin{equation}\label{fifr8a}  %
\tilde{G}(\lambda)=
\frac{2\rho_0}{\vep_2\lambda^3}
\cdot\frac{1}{[\lambda^{\alpha}+\ep\lambda+\ep^2]}\,
. \end{equation}  %
Expanding the denominator near  $\ep\lambda$, one obtains
\begin{equation}\label{fifr8b}  %
\tilde{G}(\lambda)=\frac{2\rho_0}{\vep_2\ep}
\sum_{n=0}^{\infty}\frac{(-1)^n}{\ep^n}
\sum_{m=0}^nC_n^m\frac{\ep^{2m}}{\lambda^{4+n+(n-m)\alpha}}\, ,
\end{equation}  %
where $C_n^m=\frac{\Gamma(n+1)}{\Gamma(m+1)\Gamma(n-m+1)}$.
Carrying out the inverse Laplace transform, one obtains for the
electric field
\be\label{fifr9}  %
E_r(r)=\frac{2\rho_0r}{\vep_2\ep}\sum_{n=0}^{\infty}\frac{(-1)^n}{\ep^n}
\sum_{m=0}^nC_n^m
\frac{\ep^{2m}r^{n(1-\alpha)+m\alpha}}{\Gamma[n(1-\alpha)+m\alpha+4]}\,
.\ee   %
When $\ep\ll 1$, one takes only $m=0$ in the second summation.
This yields
\be\label{fifr10}  %
E_r(r)\approx\frac{2\rho_0r}{\vep_2\ep}
\sum_{n=0}^{\infty}
\frac{[-r^{(1-\alpha)}/\ep]^n}{\Gamma[n(1-\alpha)+4]}\,
.\ee %
The sum in Eq. (\ref{fifr10}) is the definition of the
Mittag-Leffler function (see \textit{e.g.},
\cite{podlubny,klafter})
\be\label{fifr10_MLF} %
 \clE_{(1-\alpha,4)}(-r^{1-\alpha}/\ep)=
\sum_{n=0}^{\infty}
\frac{[-r^{(1-\alpha)}/\ep]^n}{\Gamma[n(1-\alpha)+4]}\,
.\ee   %
In the limit $\ep\rightarrow 0$, the asymptotic behavior of the
Mittag-Leffler function for $\frac{r^{\alpha}}{\ep}\gg 1$ has a
power law decay (see \textit{e.g}., \cite{podlubny,batmen})
\be\label{fifr10_asymptot}  %
\clE_{(\alpha,\beta)}(-cz^{\alpha})\sim
\frac{1}{\Gamma(\beta-\alpha)cz^{\alpha}}\,
, \ee  %
and we arrive at the solution (\ref{fifr3_5}) (see also
\cite{tar2005})
\be\label{fifr10_solution}  %
E_r(r)=\frac{\rho_0}{\vep_2\Gamma(\alpha+3)}r^{\alpha}\, .\ee %
Since the charge is contained inside the fractal, we also used
here that $\ep\ge 0$. This condition also ensures the transition
to a homogeneous dielectric with $\vep_1=\vep_2$ and is valid for
$\alpha=1$, as well.

In the opposite case when $\frac{r^{(1-\alpha)}}{\ep}\ll 1$, the
solution for the electric field reads
$E_r(r)\sim\frac{2r\rho_0}{\vep_2\ep}$.

%%This also corresponds to $\ep\gg 1$.
%%In this case, one
%%keeps the only term with $n=0$ in Eq. (\ref{fifr9}).

\subsection{Approximations for $\ep\ll 1$ and $\ep\gg 1$}

Note that the expansion (\ref{fifr8b}) in not always valid.
Therefore, it is convenient to use an approximation inside the
denominator in Eq. (\ref{fifr8a}). In the limiting cases $\ep\ll
1$ and $\ep\gg 1$ it can be simplified. In these cases one can
look for the solutions using an expression for the Laplace
transform for the Mittag-Leffler function for two parameters
$\clE_{(\alpha,\beta)}(-az^{\alpha})$. It reads \cite{podlubny}
\be\label{fifr16}  %
\int_0^{\infty}e^{-\lambda z}z^{\beta-1}
\clE_{(\nu,\beta)}\left(\mp az^{\nu}\right)dz=
\frac{\lambda^{\nu-\beta}}{(\lambda^{\nu}\pm a)}\, , ~~~
({\rm Re}(\nu)>|a|^{\frac{1}{\nu}})\, . \ee   %
We start by considering $\ep\ll 1$. Neglecting $\ep^2$, we
have
\be\label{fifr17}  %
\tilde{G}(\lambda)\approx\frac{2\rho_0}{\vep_2\ep\lambda^{3+\alpha}}
\cdot\frac{1}{[\lambda^{1-\alpha}+1/\ep]}\,
\ee %
Comparing Eqs. (\ref{fifr16}) and (\ref{fifr17}) one obtains
\be\label{fifr18}  %
E_r(r)\approx\frac{2\rho_0}{\vep_2\ep}r
\clE_{(1-\alpha,4)}\left(-\frac{r^{1-\alpha}}{\ep}\right)\, ,
 \ee  %
which yields the same expression for the electric field as in Eq.
(\ref{fifr10}).

In the limit $\ep\gg 1$, we neglect $\lambda^{\alpha}$ in the
denominator in Eq. (\ref{fifr8a}), which yields the following
solution for the electric field
\be\label{fifr19}  %
E_r(r)\approx\frac{2\rho_0}{\vep_2\ep}r\clE_{(1,4)}(-\ep r)\, .\ee%
Note also that the dimensionless radius can be both
$r\gg 1/\ep$ and $r\ll 1/\ep$. Therefore, for $\ep r\gg 1$  we
use the asymptotic behavior of the Mittag-Leffler
function (\ref{fifr10_asymptot}) and obtain the asymptotic behavior of
the electric field $ E_r(r)\approx \frac{\vep_2\rho_0}{\vep_1^2}$.
In the opposite case, $\ep r\ll 1$, we have
$\clE_{(1-\alpha,4)}\sim 1$, which yields
$E_r(r)\approx\frac{2\rho_0}{\vep_2\ep}r$.

\subsection{Direct Product of Cantor Sets}

In the preceding section, the random fractal was considered as a
direct product of a two dimensional sphere $S_2$ and a random
fractal $F_{\alpha}$ of the dimension $\alpha$, such that the fractal
dimension of $S_2\otimes F_{\alpha}$ is $d=2+\alpha$.
 It is worth noting, that this is only one of many
possible constructions of a random fractal \cite{falconer}.
Therefore, it is instructive to consider a more general example. Let
us consider a fractal dielectric as a product of the random Cantor
sets
$F_{\alpha}(x)\times F_{\beta}(y)\times F_{\gamma}(z)$,
such that the electric charge due to Eq. (\ref{fpe6}) is
\[Q(x,y,z)=\frac{\rho_0x^{\alpha}y^{\beta}z^{\gamma}}
{\Gamma(1+\alpha)\Gamma(1+\beta)\Gamma(1+\gamma)}\, ,~~~~x,y,z.0\]
where $\alpha+\beta+\gamma=d$ is the fractal dimension $d<3$. We
simplify the analysis by neglecting the effect of polarization
of the composite. This simplification is valid only in the case
when $\ep\ll 1$. As shown, the polarization charge is of the order
of $\ep^2$, and it can be neglected in the Maxwell equation.
Therefore, the integral Maxwell's equation reads now
\be\label{fifr12}  %
Q(x,y,z) =
(\vep_1-\vep_2)\,{}_0I_x^{\alpha}{}_0I_y^{\beta}{}_0I_z^{\gamma}
\mathbf{\nabla}\cdot\mathbf{E}(x,y,z)
+\vep_2{}_0I_x^{1}{}_0I_y^{1}{}_0I_z^{1}
\mathbf{\nabla}\cdot\mathbf{E}(x,y,z)
 \, .  \ee   %
Applying the fractional derivatives
${}_0D_x^{\alpha}{}_0D_y^{\beta}{}_0D_x^{\gamma}$ to both
sides of the equation and  denoting
$F(x,y,z)=\mathbf{\nabla}\cdot\mathbf{E}(x,y,z)$, one obtains the
Abel equation \cite{SKM,tricomi}
\be\label{fifr13_Abel}  %
\frac{\rho_0}{\vep_2}=\ep F(x,y,z)
+{}_0I_x^{1-\alpha}{}_0I_y^{1-\beta}{}_0I_z^{1-\gamma}F(x,y,z)\, .
\ee  %
Here $F(x,y,z)$ is an effective charge density. We solve this
equation by a method of consequent approximations \cite{tricomi}.
The first step is $F(x,y,z)=\frac{\rho_0}{\vep_2\ep}+F_1(x,y,z)$,
which yields the following equation for $F_1(x,y,z)$
\begin{eqnarray}\label{fifr14}  %
0=\ep F_1(x,y,z)+F_1(r) &+&
{}_0I_x^{1-\alpha}{}_0I_y^{1-\beta}{}_0I_z^{1-\gamma}F(x,y,z)
\nonumber  \\
 &+&\frac{\rho_0}{\vep_2\ep}
\frac{x^{1-\alpha}y^{1-\beta}
z^{1-\gamma}}{\Gamma(2-\alpha)\Gamma(2-\beta)\Gamma(2-\gamma)}
\end{eqnarray}%
Then, defining $F_1(x,y,z)=-\frac{\rho_0}{\vep_2\ep}
\frac{x^{1-\alpha}y^{1-\beta}z^{1-\gamma}}
{\Gamma(2-\alpha)\Gamma(2-\beta)\Gamma(2-\gamma)}+F_2(x,y,z)$ and
repeating the same procedure for finding $F_n(x,y,z)$, we have
the solution for the effective charge density in the form of the
sum
\be\label{fifr13a}    %
F(x,y,z)=\frac{\rho_0}{\vep_1-\vep_2}
\sum_{n=0}^{\infty}F_n(x,y,z)\, , \ee     %
where
\[F_n(x,y,z)=
\frac{[-x^{1-\alpha}y^{1-\beta}z^{1-\gamma}/(\vep)]^n}
{\Gamma[1+n(1-\alpha)]\Gamma[1+n(1-\beta)]
\Gamma[1+n(1-\gamma)]}\, ,\]   %
that also yields a solution for the electric field in the form of
a vector combination of $F_n(x,y,z)$ functions.

There is an essential simplification of the solutions in the
planes $x=0$, and $y=0$, and $z=0$. One has for these planes
\be\label{fifr15}    %
\nabla\cdot\mathbf{E}(x,y,z)=\frac{\rho_0}{\vep_1-\vep_2}\, .\ee  %

\section{Conclusion}
We demonstrated an application of fractional calculus for
electrostatics of composite materials in fractal geometry. The
method is based on fractional filtering of Maxwell's equation
in the framework of the Gauss theorem by means of a convolution
integral in Eq. (\ref{fpe1}).  One should recognize that the
coarse graining procedure due to the averaging over the period
$\ln b$ is important for the application of fractional calculus. We
explore the result of Ref. \cite{NIG98} with subsequent
development by splitting the composite volume into the fractal
volume $V_d\sim r^d$ with fractal dimension $d$ and the
complementary host volume $V_h$. Integrations over these
fractal volumes correspond to the convolution integrals and,
eventually, to fractional integro-differentiation. Note that
this filtering
procedure, in our case, is the averaging of the Maxwell equation, and
not of the electric field. The latter is the
solution of the obtained averaged equation. It should be noted
also that the fractal dimension alone is not enough to describe
a fractal completely. Therefore, although in the analysis above we
discuss the Cantor sets, the results are valid for any random
fractal distribution of charges, since we use the fractal
dimension only.

In this connection, we should also admit that embedding the
electric charge inside any random fractal, as in Eqs.
(\ref{rfc_1}) and (\ref{fpe6}), is based on employing the
characteristic function $\chi$ inside integrands. This integration
corresponds to the convolution integral, as the result of coarse
graining. The situation is the same for embedding the polarization
charge inside a fractal composite. Although, instead of the
characteristic function, we have its gradient $\nabla\chi$, as
shown in Appendix B, this leads to the convolution integral
(\ref{Bap6}),  as well.

One should recognize  that it is also possible that the
permittivity of the charged fractal dielectric is less than the
permittivity of the host dielectric, $\ep<0$, for example, for
biological macromolecules and polymers \cite{degennes}. In this
case the argument of the Mittag-Leffler function is positive, and
the asymptotic behavior of the electric field is absolutely different.
One finds easily that the limit $\ep\rightarrow -0$ does not
correspond to the homogeneous case with $\vep_1=\vep_2$.
Obviously $\ep=0$ is a singular point of Eq. (\ref{fifr6_b}),
since $\ep$ is the parameter at the higher derivative, and
solution (\ref{fifr18}) is singular at $\ep\rightarrow -0$ as well.
This case needs separate consideration.

We thank Liz Yodim for language editing. This research was
supported in part by the Israel Science
Foundation (ISF) and by the US-Israel Binational Science
Foundation (BSF).

\section*{Appendix A: Fractional Integration}
\def\theequation{A. \arabic{equation}}
\setcounter{equation}{0}

Extended reviews of fractional calculus can be found \textit{e.g.},
in \cite{podlubny,oldham,SKM}. Fractional integration of the order
of $\alpha$ is defined by the operator
\be\label{Ap1}  %
{}_aI_x^{\alpha}f(x)=
\frac{1}{\Gamma(\alpha)}\int_a^xf(y)(x-y)^{\alpha-1}dy\, , \ee %
where $\alpha>0,~x>a$ and  the Gamma function $\Gamma(z)$ is
defined above. Fractional derivation was developed as a
generalization of integer order derivatives and is defined as the
inverse operation to the fractional integral. Therefore, the
fractional derivative is defined as the inverse operator to
${}_aI_x^{\alpha}$, namely $
{}_aD_x^{\alpha}f(x)={}_aI_x^{-\alpha}f(x)$ and
${}_aI_x^{\alpha}={}_aD_x^{-\alpha}$. Its explicit form is
\be\label{Ap2}   %
{}_aD_x^{\alpha}f(x)=
\frac{1}{\Gamma(-\alpha)}\int_a^xf(y)(x-y)^{-1-\alpha}dy\, . \ee %
For arbitrary $\alpha>0$ this integral diverges, and as a result
of this a regularization procedure is introduced with two
alternative definitions of ${}_aD_x^{\alpha}$. For an integer $n$
defined as $n-1<\alpha<n$, one obtains the Riemann-Liouville
fractional derivative of the form
\begin{equation}\label{Ap3}   %
{}_a^{RL}D_x^{\alpha}f(x)\equiv{}_aD_x^{\alpha}f(x)
=\frac{d^n}{dx^n}{}_aI_x^{n-\alpha}f(x)\, ,
\end{equation}
and fractional derivative in the Caputo form
\begin{equation}\label{Ap4}  %
{}_a^CD_x^{\alpha}f(x)= {}_aI_x^{n-\alpha}\frac{d^n}{dx^n}f(x)\, .
\end{equation}  %
There is no constraint on the lower limit $a$. For example, when
$a=0$, one has
${}_0^{RL}D_x^{\alpha}x^{\beta}=\frac{x^{\beta-\alpha}
\Gamma(\beta+1)}{\Gamma(\beta+1-\alpha)} $. This fractional
derivation with the fixed low limit is also called the left
fractional derivative. However, one can introduce the right
fractional derivative, where the upper limit $a$ is fixed and
$a>x$. For example, the right fractional integral is
\begin{equation}\label{Ap5}
{}_xI_a^{\alpha}f(x)=\frac{1}{\Gamma(\alpha)}
\int_x^{a}(y-x)^{\alpha-1}f(y)dy\, .
\end{equation}  %
Another important property is
\be\label{Ap6}  %
D^{\alpha}D^{\beta}=D^{\alpha+\beta}~~~ \mbox{and}~~~~
D^{\alpha}I^{\beta}=I^{\beta-\alpha}\, , \ee   %
where other indexes are omitted for brevity. We also use here a
convolution rule for the Laplace transform for $0<\alpha<1$
\begin{equation}\label{mt3}
\clL[{}I_x^{\alpha}f(x)]=s^{-\alpha}\tilde{f}(s)\, .
\end{equation}
Note that in physical applications a treatment of the Caputo
fractional derivative by the Laplace transform is more convenient
than the Riemann-Liouville one. In the first case, one has to
define $0\le n<\alpha$ integer derivative at the initial point,
while for the second case one defines  $n$ Riemann-Liouville
fractional derivative as initial conditions. Nevertheless, the use
of either Caputo or Riemann-Liouville fractional derivatives is
equivalent for initial value problems and leads to the same
results \cite{klafter,Metzler}.

\section*{Appendix B}
\def\theequation{B. \arabic{equation}}
\setcounter{equation}{0}

The last term in the l.h.s. of Eq. (\ref{fifr6_a}) corresponds to
the induced charge, or polarization charge on the fractal, since
the electric field is a discontinuous function \cite{LLVIII}. We
also note that, while the notation $\nabla_r$ is the divergence in the
Gauss theorem $\nabla_r=\frac{1}{r^2}\frac{\prt}{\prt r}r^2$, it
is the gradient in the last term
$\nabla_r\chi(r)=\frac{\prt\chi(r)}{\prt r}$. Let us consider the
$N$th step of the fractal construction. It is a union of disjoint
intervals $\Delta_N$. In the limiting case one obtains
$F_{\alpha}=\lim_{N\to\infty}\bigcup\Delta_N$. Therefore, $\forall
r_j\in F_{\alpha}$ the characteristic function on every
closed interval $[r_j,r_j+\Delta_N]$, is $\chi(\Delta_N)=
\Theta(r-r_j)-\Theta(r-r_j-\Delta_N)$. Differentiation of the
characteristic function on the intervals yields
\be\label{Bap1}  %
\frac{\prt}{\prt r}\chi(\Delta_N) =
\delta(r-r_j)-\delta(r-r_j-\Delta_N)\, .\ee  %
Therefore, we have for any interval $\Delta_N$ and at $r=r_j$
\be\label{Bap2} %
{}_rI_{r+\Delta_N}^1E(r)r^2\nabla_r\chi(r) =
E(r)r^2-E(r+\Delta_N)(r+\Delta_N)^2\, . \ee %
This expression is not zero in the limit $\Delta_N\rightarrow 0$,
since $\vep_2E(r-0)=\vep_1E(r+0)$ and
$\vep_1E(r+\Delta_N-0)=\vep_2E(r+\Delta_N+0)$ \cite{LLVIII}. For
brevity's sake the index $r$ for the electric field is omitted
$E_r(r)\equiv E(r)$.
We make a shift between points $r_j-0$ and $r_j+\Delta_n-0$.
We present the result of integration in Eq. (\ref{Bap2})
in the form
\begin{eqnarray*}
&r^2[E(r-0)-E(r+\Delta_N-0)]+{\rm O}(\Delta_N)= \nonumber \\
&r^2[E(r-0)-E(r+0)+E(r+0)-E(r+\Delta_N-0)]+{\rm O}(\Delta_N)\, .
\end{eqnarray*}
Since inside the interval $\Delta_N$ the electric
field is continuous, we have $E(r+\Delta_N-0)\approx
E(r+0)+E'(r+0)\Delta_N$. Now the boundary conditions for the
electric field at $r_j$ can be taken into
account: $\vep_2E(r_j-0)=\vep_1E(r_j+0)$. This, finally, yields
\be\label{Bap3} %
\lim_{\Delta_N\to 0}{}_rI_{r+\Delta_N}^1E(r)r^2\nabla_r\chi(r)=
\frac{\vep_1-\vep_2}{\vep_2}r^2E(r)~~~ \forall r\in F_{\alpha}\, .  \ee %

We have from Eq. (\ref{fifr6_me}) that
${}_0I_r^1E_r(r)r^2\nabla_r\chi(r)$ is a superposition of random
values, namely
\be\label{Bap4}  %
{}_0I_r^1E_r(r)r^2\nabla_r\chi(r)=
\frac{\vep_1-\vep_2}{\vep_2}\sum_{r_j\in
F_{\alpha}}r_j^2E_r(r_j)\, . \ee %
This can be approximately considered as an ``average'' value
${}_0I_r^1\chi(r)E_r(r)r^2\rightarrow{}_0I_r^{\alpha}E_r(r)r^2$.
Indeed, let us present the sum in the form
\be\label{Bap5}  %
\sum_{r_j\in F_{\alpha}}r_j^2E_r(r_j)=\sum_{r_j\in
F_{\alpha}}\int_0^rr'^2E_r(r')\delta(r'-r_j)\, . \ee %\, .
Hence, one obtains the integration of the electric field with
the fractal density $\sum_{r_j\in F_{\alpha}} \delta(r'-r_j)$. This
corresponds to the fractal volume $r^{\alpha}$. The next
step is to consider the integration in Eq. (\ref{Bap5}) as
the convolution
integral with the kernel $(r-r')^{\alpha-1}$ \cite{Ren}.
Eventually, we obtain for the polarization charge term in Eq.
(\ref{fifr6_a})
\be\label{Bap6}  %
{}_0I_r^1E_r(r)r^2\nabla_r\chi(r)\rightarrow\frac{\vep_1-\vep_2}{\vep_2}
{}_0I_r^{\alpha}E_r(r)r^2\, .\ee %
Finally, we note that the shift in the intervals $\Delta_N$ is
taken between point $r_j-0$ and $r_j+\Delta_N)$.
One can also make the shift between points $r_j+0$
and $r_j+\Delta_N+0$. The difference
between electric fields at the points yields
$[1-\frac{\vep_1}{\vep_2}]E(r_j)$. Taking into account that
this difference is opposite to the external normal, we change
the sign and obtain the same expression as in Eq. (\ref{Bap3}).

\end{document}